# Engineering Transport in Manganites by Tuning Local Non-Stoichiometry in Grain Boundaries

*F. Chiabrera, I. Garbayo, L. López-Conesa, G. Martín, A. Ruiz-Caridad, M. Walls, L. Ruiz-González, A. Kordatos, M. Núñez, A. Morata, S. Estradé, A. Chroneos, F. Peiró, A. Tarancón\**

F. Chiabrera, Dr. I. Garbayo, Dr. M. Núñez, Dr. A. Morata, Prof. Dr. A. Tarancón
Department of Advanced Materials for Energy, Catalonia Institute for Energy Research (IREC), Jardí de les Dones de Negre 1, Planta 2, 08930 Sant Adrià de Besòs (Barcelona), Spain.

Dr. L. López-Conesa, G. Martín, A. Ruiz-Caridad, Dr. S. Estradé, Prof. Dr. F. Peiró
Department of Electronics and Biomedical Engineering, University of Barcelona, C. de Martí i Franquès 1, 08028 Barcelona, Spain.

Dr. L. López-Conesa, G. Martín, A. Ruiz-Caridad, Dr. S. Estradé, Prof. Dr. F. Peiró
Institute of Nanoscience and Nanotechnology, University of Barcelona, 08028 Barcelona, Spain.

Dr. L. López-Conesa
TEM-MAT Unit, Scientific and Technological Centers of the University of Barcelona (CCiTUB), C. de Lluís Solé i Sabarís 1, 08028 Barcelona, Spain.

M. Walls
Laboratoire de Physique des Solides Bldg. 510, CNRS, Université Paris-Sud, Université Paris-Saclay, Orsay Cedex, 91405 France.

Prof. Dr. L. Ruiz-González
Departamento de Química Inorgánica, Facultad de CC. Químicas, Universidad Complutense de Madrid, 28040 Madrid, Spain.

A. Kordatos, Prof. Dr. A. Chroneos
Faculty of Engineering, Environment and Computing, Coventry University, Priory Street, Coventry CV1 5FB, United Kingdom.

Prof. Dr. A. Chroneos
Department of Materials, Imperial College London, London SW7 2AZ, United Kingdom.

Prof. Dr. A. Tarancón
ICREA, Passeig Lluís Companys 23, 08010 Barcelona, Spain.

E-mail: atarancon@irec.cat

Interface-dominated materials such as nanocrystalline thin films have emerged as an enthralling class of materials able to engineer functional properties of transition metal oxides widely used in energy and information technologies. In particular, it has been proved that strain-induced defects in grain boundaries of manganites deeply impact their functional properties by boosting their oxygen mass transport while abating their electronic and magnetic order. In this work, the origin of these dramatic changes is correlated for the first time with strong modifications of the anionic and cationic composition in the vicinity of strained grain boundary regions. We are also able to alter the grain boundary composition by tuning the overall cationic content in the films, which represents a new and powerful tool, beyond the classical space charge layer effect, for engineering electronic and mass transport properties of metal oxide thin films useful for a collection of relevant solid state devices.

Perovskite manganites with general formula $RE_{1-x}B_xMnO_{3\pm\delta}$ (where RE stands for a trivalent Rare-Earth element and B for a divalent alkaline ion) have been extensively investigated for their wide variety of intriguing properties, such as oxygen electrocatalysis in solid oxide fuel cell (SOFC),[1,2] colossal magnetoresistivity,[3,4] resistive switching[5,6] and magnetocaloric effect.[7,8] Classically, engineering the electro-chemo-mechanical properties in such materials (either bulk or thin films) was reached by the selection of the RE element and/or by doping the A-site with aliovalent atoms. However, in the last years the study of interface-dominated thin films has opened new horizons in the understanding and control of functional transition-metal oxides. For instance, the control of the strain imposed by a substrate in epitaxial thin films was demonstrated to have a large impact on many different aspects, such as the electronic and magnetic properties,[9,10] or the oxygen surface exchange rate.[11] In addition to the study of pure strain effects, also structural defects, such as dislocations or grain boundaries (GBs), have been shown to strongly influence the overall properties of manganites.[8,12–16]

It was recently discovered that GBs in $La_{0.8}Sr_{0.2}MnO_{3\pm\delta}$ (LSM) thin films are responsible for a substantial enhancement of oxygen diffusion and oxygen reduction reactions (ORR),[13,14,17] contrary to what is determined in many oxides where the GBs are known to hinder the oxygen mobility.[18,19] The improvement of oxygen mass transport properties is able to completely change the nature of the material converting the mainly electronic LSM into a good mixed ionic-electronic conductor (MIEC). Moreover, it was also determined that dislocations in epitaxial compressive thin films can accelerate oxygen diffusion, suggesting an interesting similarity with the behaviour of the GBs.[16] This substantial difference between LSM and other oxides is at the centre of an important debate within the scientific community, since the comprehension of the phenomena might open new horizons to take advantage of GB peculiarities in several families of solid state electrochemical devices.

Alternatively, GBs are also known to deeply impact the electronic and magnetic properties of manganites, generally giving rise to a highly insulating and magnetically disordered region,[12,20,21] which can strongly affect, among others, the performance of magnetic cooling devices[8] or magnetic memories.[22] Since the electronic transport properties of manganites are determined by the chemical and structural order of Mn,[3,23] phenomena occurring at the GB level such as lattice strain, cation deficiency and variable oxygen concentration will certainly have an impact on the electronic behaviour.

In this direction, the present study correlates structural and local compositional changes in the vicinity of GBs with the changes observed in the electrochemical and electronic behaviour of LSM thin films, revealing the origin of both their superior ORR catalysis and their deleterious higher resistivity. Ultimately, we prove that the modification of the local composition at the GB level allows tuning these electrochemical, electronic and magnetic properties thus proposing a new and effective tool for engineering the functionality of these, and other technologically relevant materials.

To understand the correlation between the superior performance of dense LSM layers and their local structure, we first prepared Mn-deficient polycrystalline films by means of Pulsed Laser Deposition (PLD) on top of yttria stabilized zirconia (YSZ)-based substrates (YSZ/$Si_3N_4$/$SiO_2$/Si), using a proper selection of deposition parameters.[24–26] LSM films showed a fully dense polycrystalline structure, with well-defined nanometric columnar grains of tens of nanometers in diameter (**Figure 1a** and **1b**) and a Mn/(La+Sr) ratio of B/A = 0.85 ± 0.02, as measured by Wavelength Dispersive Spectroscopy (WDS). High Resolution Transmission Electron Microscopy (HR-TEM) images of the cross section (**Figure 1c**) revealed the good crystallinity of the LSM grains and the absence of secondary phases neither at the GB nor at the core of the columnar-like grains. The most relevant feature of the grain boundaries, observed after Fourier filtering of HRTEM images, is the presence of a high concentration of ordered lattice defects with a periodicity of ∼1nm (**Figure 1d**, see also **section S.1.** of the supplementary information). Higher magnification images show the existence of a regular dense array of dislocations, consisting of a missing atomic plane one out every five (**Figure 1e**).

This dislocation network imposes an elastic stress field that propagates for several nanometers inside the grains. Eventually, due to the short distance between the dislocations, the strain field overlaps determining a strained region far beyond the grain boundary core.[27]

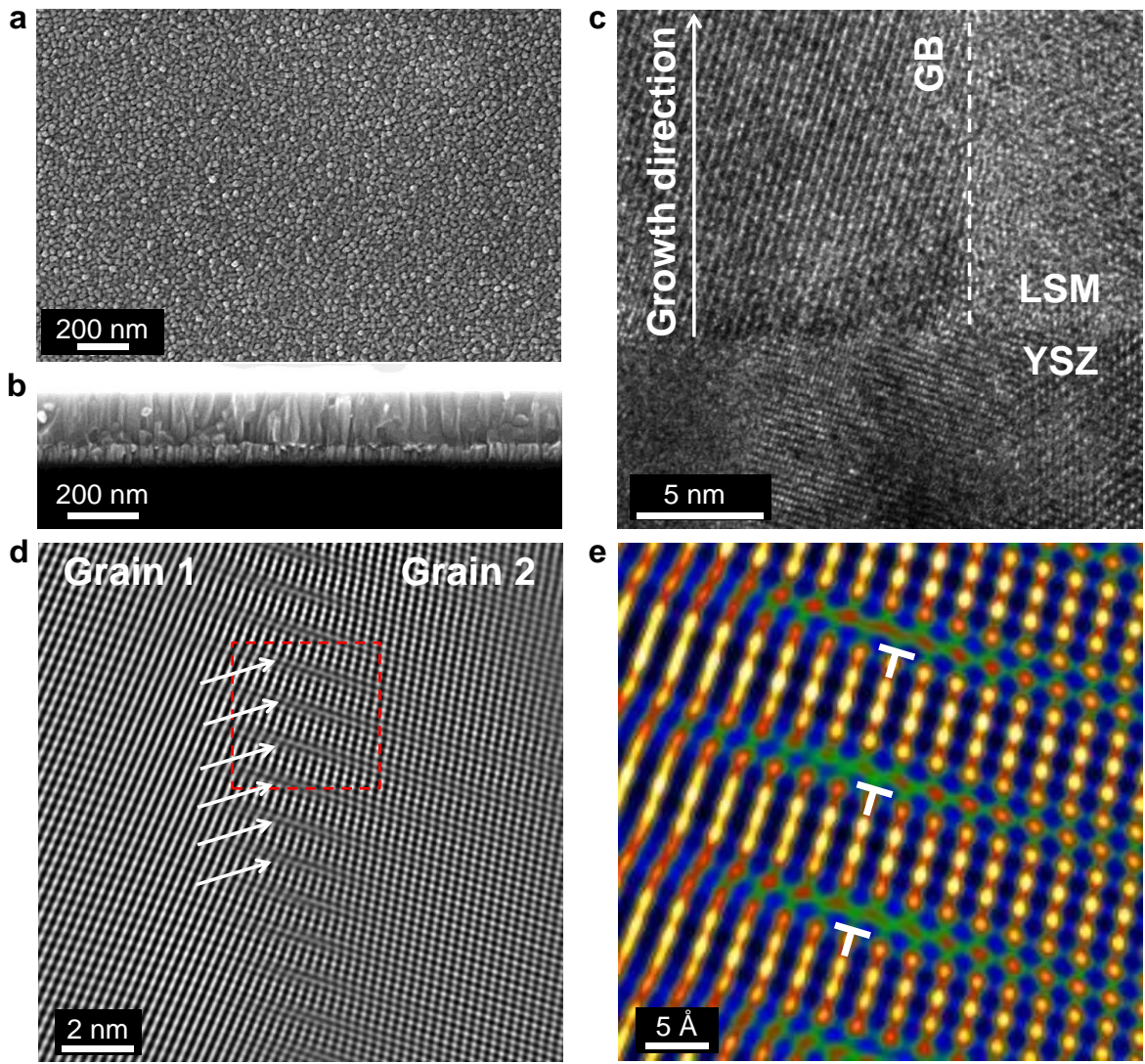

**Figure 1. a,** SEM top-view and **b,** cross-section of the fully-dense and columnar LSM/YSZ bi-layer grown on the top of $Si_3N_4/SiO_2/Si$. **c.** HRTEM image of YSZ/LSM interface showing one grain boundary (highlighted by a white dashed line) oriented in a direction perpendicular respect to the substrate. **d,** Fourier filtered image of a LSM grain boundary consisting of a dense pattern of dislocations repeating every ~ 1 nm. **e,** coloured enlargement of the red dotted square shown in **d**, in which 3 dislocations are highlighted by white "T" symbols.

The effect of these highly defective GBs on the chemical composition was analysed by high angle annular dark-field (HAADF) and electron energy-loss spectroscopy (EELS). First, HAADF (**Figure 2a**) clearly showed a strong contrast associated to the defective region, indicating a local variation in the atomic number likely related to deviations from the bulk grain stoichiometry. More precisely, EELS analysis across and along the grain boundary (**Figure 2b and 2d**, respectively) revealed this significant variation in the atomic composition. Across the interface, **Figure 2b**, we observed an increase of the concentration of Sr and La coupled to a decrease in Mn and O, common to all GBs analysed (see also **Figure S4** in the Supplementary Information). Importantly, despite the stoichiometry deviation, a constant Mn $L_{23}$ ratio of *ca.* 2.5 (corresponding to a Mn oxidation state of ~3.4)[28] was measured across the GB (**Figure S5**). Moreover, the EELS measurements along the GB interface, **Figure 2c**, yielded analogous compositional variations with complementary La and Mn profiles when crossing the local strain fields created by the dislocations (**Figure 2d**). This latter observation suggests that the strained regions generated by the dislocations are strongly related with the local variation of the stoichiometry. Although the observed oxygen deficiency is fully compatible with the well-known reducing nature of dislocations,[18,29–31] the presence of a cation rearrangement (combined with the constant oxidation state of Mn) indicates a more complex charge compensation mechanism than the classical space charge layer formation. The presence of large concentration of manganese vacancies in the bulk (B/A = 0.85 ± 0.02) suggests an interdependence between the grain interior and the GBs, opening the way for tuning the local non-stoichiometry at the GB level by controlling the overall cationic content in the films.

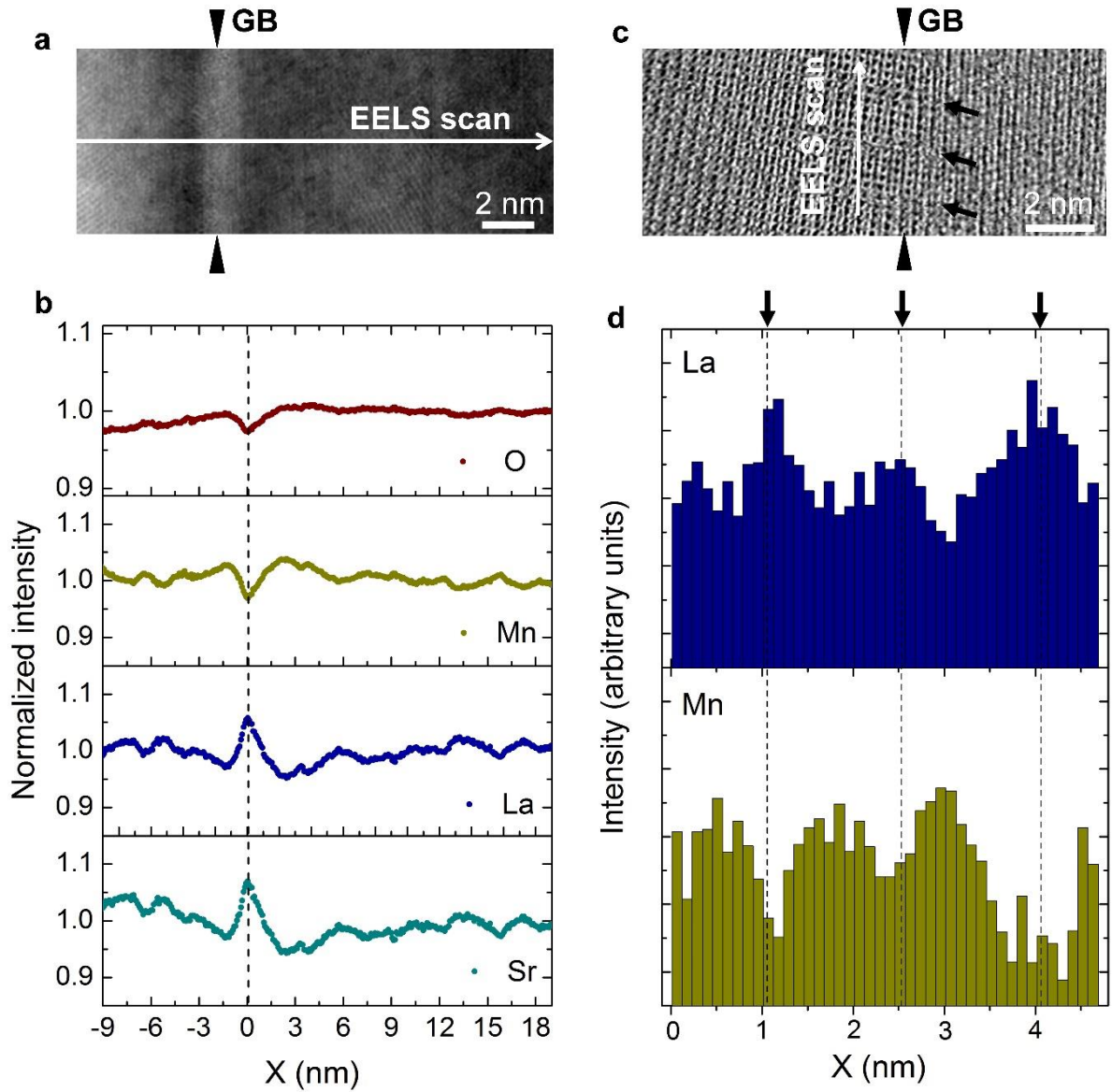

**Figure 2. a,** HAADF image of the GB, showing a different contrast at the interface. **b,** chemical composition obtained from the EELS scan across the GB (white arrow in **a**). Integrated intensity signals were normalized to the grain (bulk) values for each element. A significant compositional variation is visible near the GB interface. **c,** Bright field HRTEM figure of a grain boundary showing atomic fringes at the interface (indicated by black arrows) that agree with the dislocations pattern observed in **Figure 1.c. d,** La and Mn variation extracted from EELS scan along the GB (white arrow in **c**). A simultaneous increase of La and decrease of Mn concentration is observed at specific locations that correspond well with the fringe pattern observed, indicating that the chemical variations occurs in the proximity of the dislocations.

We therefore investigated the possibility of tuning the ionic composition in the GBs and thus, the functional properties of the LSM films. For this purpose, combinatorial pulsed laser deposition (c-PLD) was employed to generate a thin film with a continuous map of compositions $La_{0.8}Sr_{0.2}Mn_yO_{3\pm\delta}$ ($LSM_y$, y = 0.85-1.2). LSM and $Mn_3O_4$ targets were used as parent compounds, as shown in **Figure 3a** (see **section S.3.1.** for further details on the fabrication process). High quality and dense polycrystalline thin films with variable content of Mn were grown on different substrates, as confirmed by Scanning Electron Microscopy (SEM) and Energy Dispersive X-ray (EDX) analysis (**section S.3.2.** includes a complete structural characterization of the combinatorial film). Importantly, the structural characterization of the $LSM_y$ film carried out by X-Ray Diffraction (XRD) confirmed the absence of secondary phases in the whole Mn/(La+Sr) range under study and independently of the substrate, indicating that the Mn was effectively introduced in the perovskite structure (**Figure 3b**). The expected continuous increase of the pseudo-cubic lattice parameter $c_{pc}$ in the Mn-deficient part together with negligible variations in the La-deficient part was obtained (**Figure 3c**).[32,33] Analogously to the study of Mn-deficient LSM films, HAADF and EELS analysis of the $LSM_y$ layers was carried out by TEM. Here, a clear impact of the change in the bulk stoichiometry on the local cation composition at the grain boundary level was observed, see representative GBs for each composition in **Figure 3d-k**. The analysis reflects that the GBs suddenly pass from a Mn-deficient state to a La-deficient with the overall Mn content, presumably changing the defect chemistry and giving rise to two different types of GBs, with opposite cation depletion (for a discussion on the quantitative variations we refer to section S.3.3 in the Supplementary Information). Notably, oxygen deficiency was found in all the GBs, independently on the Mn content (see **Figure S14** in the Supplementary Information). In order to assess the impact of the local non-stoichiometry of the GBs on the transport properties of the $LSM_y$ films, we evaluated for the whole series of compositions both the in-plane electrical conductivity (measured on insulating substrates, Sapphire (0001) and $YSZ/Si_3N_4/SiO_2/Si$) and the electrochemical properties (on an oxygen ion conducting electrolyte, YSZ (001)), see **Figures 4a and 4b** respectively. First, we focus on the changes observed in the electrical properties, **Figure 4c**. Here, we show the evolution of the electronic resistivity as a function of temperature (T= 85-575 K) for the polycrystalline $LSM_y$ series on a sapphire substrate. Note that, for comparison, the plot also includes results obtained for a 60 nm-thick epitaxial LSM film, grown on an NdGaO$_3$ (110) substrate under the same deposition conditions than the polycrystalline LSM$_{0.85}$ (see **section S.2.** for a complete characterization). This epitaxial film is expected to behave mostly like the grain interior of polycrystalline LSM$_{0.85}$ and is taken as a reference to ensure that the effects observed are associated to the GBs. Indeed, the polycrystalline and epitaxial LSM$_{0.85}$ thin films present the same bulk cationic composition, Mn oxidation state and, consequently, oxygen content.[34] As shown in **Figure 4c**, the epitaxial LSM$_{0.85}$ film (blue open symbols) indeed presents a metal-insulator (M-I) transition typical of bulk stoichiometric manganites. Although a suppression of M-I transition was previously reported for Mn deficient compositions,[24,35,36] the absence of extended structural defects (GBs) in the epitaxial films preserves the long range electronic order, allowing the onset of metal behaviour. Opposed to that, the polycrystalline counterpart (LSM$_{0.85}$) presents at least one order of magnitude higher resistance together with the suppression of the M-I transition. Moreover, it is important to note that no changes in the electrical resistivity were observed after annealing at temperatures as high as 923 K for 5 hours, suggesting a strong equilibrium of the GB (see **Figure S17** in Supplementary Information). Remarkably, as shown in **Figure 4c**, progressively increasing the B/A ratio leads to a continuous decrease of the resistivity recovering the low temperature M-I transition for Mn/(La+Sr) > 0.92 (complementary results were obtained for the YSZ/Si substrate as shown in the supplementary information, **Figure S18**).

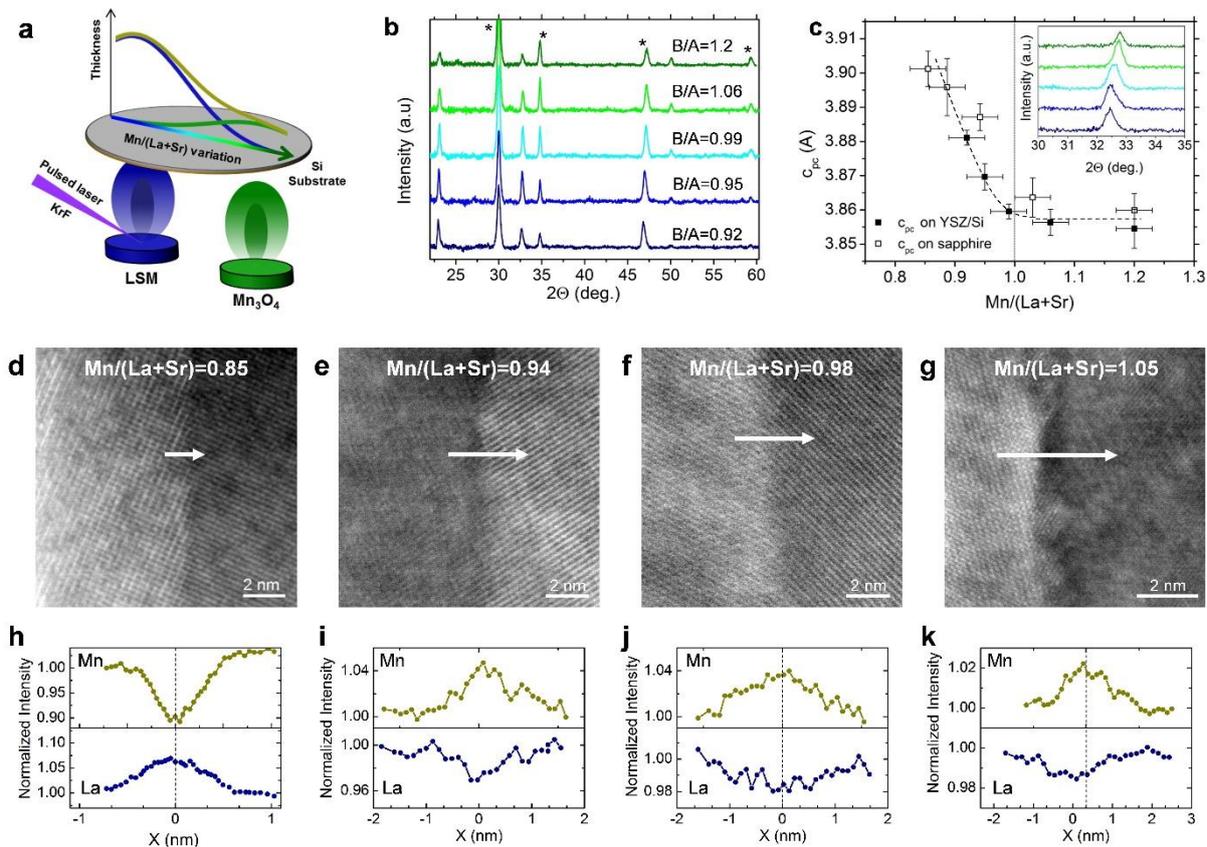

**Figure 3. a**. Sketch of the combinatorial approach used for varying B/A ratio. **b**. XRD of LSM$_y$ samples deposited on YSZ/Si for different B/A ratio. Black asterisk marks the YSZ diffraction peaks. **c.** Evolution of pseudo-cubic lattice parameter obtained by XRD for different Mn content. The inset shows the shift of the 110 peak in the samples deposited on sapphire (0001) **d-g**, Annular dark field images of grain boundaries in samples with different B/A ratio and **h-k** corresponding La and Mn variation extracted from EELS scans. A sudden decrease of La and a progressive increase of Mn are obtained in the GBs changing the overall B/A ratio in the thin films.

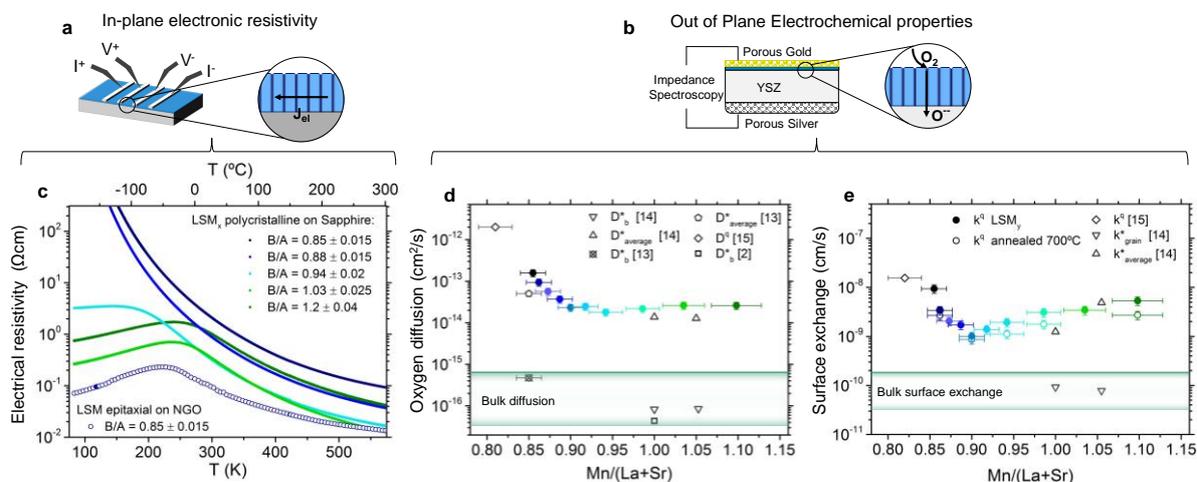

**Figure 4. a.** Sketch of the in-plane electrical measurement and c. resistivity obtained as a function of temperature for different B/A ratio. Results obtained for the epitaxial LSM$_{0.85}$ are reported for comparison **b.** Scheme of the electrochemical measurements. The columnar structures of the films offer a preferential ionic pathway along the grain boundaries. Oxygen

diffusion (**d**) and surface exchange coefficient (**e**), extracted from the impedance at 650ºC in air for different B/A ratio. Diffusion coefficients measured in SIMS experiments by Saranya et al.,[13] Navickas et al.,[14] De Souza et al.[2] and in EIS by Chiabrera et al.[15] are reported for comparison. Stoichiometry of films from references [13,15] were measured within the frame of this work.

This M-I transition in manganites has been traditionally interpreted as a Jahn-Teller (JT) mediated double-exchange mechanism, consisting of the transfer of one electron between two ferromagnetic spin-aligned Mn ions through the oxygen orbital.[37,38] In the Mn-deficient LSM GBs (see **Figure 2**), the simultaneous decrease of O and Mn is expected to strongly impede the metallic behaviour due to the interruption of the Mn-O-Mn chains and the formation of a highly insulating zone of the width of the local non-stoichiometry. The recovery of the M-I transition in the polycrystalline thin films well correlates with the change of GB composition observed increasing the overall B/A ratio (see **Figure 3**). In these Mn-rich GBs, the decrease of La is not supposed to affect the electronic transport, while the progressive increase of Mn helps to restore the metallic long-range order, gradually improving the metallic behaviour. It is worth mentioning that also an enhancement of the bulk conductivity was measured in the epitaxial thin films increasing the Mn content (see **Figure S19** in Supplementary Information).[34] Nevertheless, the conductivity of the polycrystalline thin films is orders of magnitude lower than the corresponding epitaxial one, meaning that the GB contribution is always dominating the macroscopic electronic transport.

Complementary to the electrical measurements, we carried out the electrochemical characterization of the $LSM_y$ series as part of $LSM_y$/YSZ/Ag cells, where YSZ is a pure ionic conductor acting as an electrolyte and the silver is playing the role of the backside electrode. The study of this type of MIEC-based cells by Electrochemical Impedance Spectroscopy (EIS) allows the determination of their oxygen mass transport properties,[1] i.e. the oxygen diffusion coefficient, $D^*$, and the surface exchange coefficient, $k^*$ (see section **S.3.5.** for a detailed explanation). The evolution of the oxygen diffusion and surface exchange coefficients obtained by EIS analysis in air at 923K are presented in **Figure 4d** and **e**, respectively. First, a substantial increase of the diffusivity (above two orders of magnitude) with respect to the reported bulk values is observed in all the samples. This confirms the existence of an oxygen fast diffusion pathway along the GBs, independently of the Mn/(La+Sr) ratio, consistent with the oxygen deficiency observed in all the GB types (**Figure S14**). On top of that, Mn-deficient layers with Mn/(La+Sr) < 0.9 present an additional increase up to two orders of magnitude, for a total of four, in the diffusivity. On one side, these results may be behind the discrepancies observed by different authors in the past,[13–15] while on the other they show for the first time the possibility of tuning the diffusivity by controlling the content of Mn. Regarding the surface exchange (**Figure 4e**), an analogous trend to the oxygen diffusion is observed in the Mn deficient GBs. The strong coupling of the two oxygen parameters, along with the observed oxygen depletion in the GBs, corroborate the hypothesis that the accumulation of oxygen vacancies is at the origin of the large improvement of the Oxygen Reduction Reactions (ORR) reported for manganite thin films by different authors.[17] This conclusion is also supported by theoretical works from Choi *et al.* [39] and Mastrikov *et al.*,[40] where the scarce concentration of oxygen vacancies was found as the limiting factor for oxygen incorporation in bulk LSM. The strong implications of our observation contribute to the general understanding of the ORR mechanism for metal transition based perovskites, which is still under debate.[41–43] In this direction, a traditional gap of knowledge regarding the unexpected excellent behaviour of a mainly electronic conductor such as LSM bulk is overcome by considering the superior interface performance proved, when the material is not limited by the availability of oxygen vacancies.

With these results we unambiguously show that the origin of the deleterious effects on the electrical behaviour and the superior oxygen mass transport properties rely on the evolution of the nature of the GB, which is dominated by the cation and anion local non-stoichiometry. A transition in the nature of GBs occurs at Mn/(La+Sr) ~ 0.9 in LSM, likely associated to a change in the local defect chemistry. Above this value, slightly Mn-deficient and Mn-rich films are characterized by a significant oxygen vacancies concentration ($V_O^{\cdot\cdot}$), arising from the presence of highly reducing dislocations,[31] compensated by lanthanum vacancies ($V_{La}'''$) and the formation of naturally occurring antisite defects ($Mn_{La}^\times$),[44] which is fully compatible with our HRTEM-EELS observations. More interestingly, for highly Mn-deficient layers (Mn/(La+Sr) < 0.9), an increasing oxygen diffusivity is observed due to a further increase of the oxygen vacancies concentration, which seems to be strongly interrelated to the decrease in Mn content and the presence of compensating negative defects such as Mn vacancies ($V_{Mn}'''$) (**Figure 2**). As previously mentioned, this decrease of Mn at the GB level is complemented by an increase of the La and Sr content. This, together with the absence of secondary phases (such as the La-rich Ruddlesden-Popper phase) in the GBs, suggests the formation of the $La_{Mn}^\times$ antisite defect. However, this type of defect is hardly formed in strain-free bulk LSM due to the much bigger ionic radius of La$^{3+}$ compared to Mn$^{3+}$/Mn$^{4+}$.[45] Density Functional Theory (DFT) simulations were carried out in this work (see details in the experimental section) to evaluate the formation of these unusual $La_{Mn}^\times$ defects in combination with other defects and constrains typically present at the LSM GBs, i.e. strain, the presence of oxygen vacancies and Sr dopants. The formation energies of different defect clusters under variable strain states ($\varepsilon$) were calculated to find the most stable defect configurations in the presence of Mn vacancies (see **Table 1**). Several conclusions arise from this analysis: first, the formation of the oversized $La_{Mn}^\times$ defect is more favourable in the vicinity of the smaller Sr cations ($Sr_{La}'$); second, the formation energy is even lower in the presence of one oxygen vacancy, $V_O^{\cdot\cdot}$ (additional Sr dopants in the cluster do not reduce this value); third, no significant variation was obtained when applying strain in the presence of Sr doping atoms and/or oxygen vacancies. All in all, this indicates that the driving energy for the antisite defect formation in GBs relies on a complex association of defects ($V_{Mn}''', Sr_{La}'$ and $V_O^{\cdot\cdot}$) rather than a classical space charge mechanism.

**Table 1.** Formation energy of $La_{Mn}^{\times}$ defect calculated by DFT in different configurations.

| Formation energy (eV) | | Without $V_O^{\cdot\cdot}$ | | | With one $V_O^{\cdot\cdot}$ | | |
|---|---|---|---|---|---|---|---|
| | | $\varepsilon = 0$ | $\varepsilon = 1\%$ | $\varepsilon = 2\%$ | $\varepsilon = 0$ | $\varepsilon = 1\%$ | $\varepsilon = 2\%$ |
| # $Sr'_{La}$ | 0 | 5.25 | - | - | 2.29 | - | - |
| | 1 | 3.36 | 3.34 | 3.43 | 2.61 | 2.58 | 2.66 |
| | 3 | 2.96 | 2.48 | 2.13 | 2.32 | 2.35 | 2.3 |

It is important to note that local cation non-stoichiometry in manganites was also observed in other type of interfaces, e.g. dislocations in epitaxial La$_{0.7}$Sr$_{0.3}$MnO$_3$ on LaAlO$_3$[46,47] or purely strained La$_{0.7}$Sr$_{0.3}$MnO$_3$/SrTiO$_3$ hetero-interfaces.[48–50] This suggests that the compensation mechanism observed in this work can be generally extrapolated to other strained and oxygen deficient interfaces. The natural presence of cationic vacancies in manganites, even in the nominally stoichiometric material,[51] is most likely the main driver of this unusual elemental rearrangement. The GB engineering shown in this work could be therefore extended to other interfaces and, more interestingly, to other perovskites, opening the way for new strategies to control the transport properties of materials by tuning the local non-stoichiometry.

In summary, we revealed the origin of the superior mass transport properties of interface-dominated LSM thin films by directly observing a strain-related accumulation of oxygen vacancies in the GBs that defines high diffusion pathways. We found a strong correlation between these changes and the local cation non-stoichiometry observed at the GB level, which we were able to actively modify for completely change the nature of the interface. Therefore, we present here a new tool for tailoring the transport properties of technologically relevant interfaces such as grain boundaries (beyond the traditional space charge layer engineering) suitable for direct application in a collection of solid state devices based on mixed ionic-electronic conductors, including thin film based solid oxide fuel cells, resistive switching devices, magnetoresistance or supercapacitors.

**Experimental Section**

*Films.* La$_{0.8}$Sr$_{0.2}$Mn$_{1-y}$O$_{3\pm\delta}$ (LSM) / Yttria-stabilized Zirconia (YSZ) polycrystalline heterostructures (100nm LSM / 100 nm YSZ) were grown by PLD on Si (100) substrates, electrically and stress passivated by 300nm Si$_3$N$_4$ / 100 nm SiO$_2$ bilayers. The PLD system used is a large-area system from PVD Products (PLD-5000), equipped with a KrF-248 nm excimer laser from Lambda Physics (COMPex PRO 205). Commercial targets with nominal composition La$_{0.8}$Sr$_{0.2}$MnO$_3$ (LSM) and 3 mol-% Y$_2$O$_3$ stabilized ZrO$_2$ (3YSZ) were used for the thin films deposition. The films were deposited with an energy fluency of 1 J cm$^{-2}$ per pulse at a frequency of 10 Hz. The substrate was kept at 700ºC, in an oxygen partial pressure of 0.026 mbar during the deposition and the substrate-target distance was set to 95mm. Epitaxial LSM films were deposited under the same conditions of the polycrystalline films on NdGaO$_3$ (110) substrates. Combinatorial pulsed Laser deposition (c-PLD) technique was used to grow LSM$_y$ films with variable Mn/(La+Sr) ratio. Commercial targets of LSM and Mn$_3$O$_4$ were used as parent compounds. The deposition conditions used were the same as for the single polycrystalline and epitaxial LSM. LSM$_y$ combinatorial samples were grown on YSZ/Si 4" wafers and 5mm x 10mm Sapphire (0001) substrates (for the in-plane electrical measurements) and on 3mm x 3mm YSZ (001) single crystal (for the electrochemical measurements). More details on the combinatorial deposition technique can be found in **section S.3.1.** of the supplementary information. The nominal composition of the films was measured by Wavelength Dispersive Spectroscopy (WDS) (model- Jeol JXA-8230) technique and by Energy Dispersive X-ray Analysis. The films were characterized by SEM in a ZEISS AURIGA equipment and by XRD in a four-angle diffractometer with a Cu Kα radiation source (X'Pert MRD-Panalytical) and in a Bruker D8 Advance diffractometer system.

*Electrical characterization.* LSM thin films were analysed in 4-points in-plane electrical measurements in a temperature-controlled Linkam probe station, between 83K and 573K (more details can be found in **section S.3.3.1** of the supplementary information).

*Electrochemical characterization.* Electrochemical impedance spectroscopy (EIS) was performed on the LSM$_y$ combinatorial dense samples deposited on YSZ single crystals with a Novocontrol Alpha-A analyzer. Porous silver paste was painted as backside electrode due to its excellent oxygen electrochemical properties. Porous gold paste was applied on the top of the LSM$_y$ films for assuring a homogenous current collection. The frequency range chosen was $10^6$-0.05 Hz and an AC voltage with amplitude 0.05 V was applied. The experiments were performed in a ProboStat test station (NorECs) placed inside a vertical furnace. The temperature was set to 650ºC and synthetic air was flown into the system.

TEM. HRTEM images were acquired in a JEOL J2010F microscope operated at 200 kV accelerating voltage. EELS experiments were carried out in a JEOL ARM 200cF microscope operated at 200 kV, a FEI TITAN Low Base operated at 300 kV and a Nion UltraSTEM operated at 100 kV.

*DFT simulations.* The calculations were performed using the plane wave DFT code CASTEP.[52,53] The exchange and correlation interactions were modelled with the corrected density functional of Perdew, Burke and Ernzerhof (PBE)[54] within the generalized gradient approximation (GGA) and ultrasoft pseudopotentials.[55] Spin polarized calculations with the Hubbard U contribution were used to account for the strong Coulombic interaction of the localised electrons of the 3d orbitals of Mn (the U value was set to 4 eV). The plane wave basis set was set to a cut-off of 400 eV, in conjunction with a $2 \times 3 \times 2$ Monkhorst-Pack (MP)[56] k-point grid and an 80-atomic site supercell with lattice parameters of a = c = 11.1 Å, b = 7.73 Å and supercell volume of 954.9 Å$^3$. LMO / LSM is an orthorombic perovskite within the (Pnma) space group (space group number 62). All supercells were geometry optimized and fully relaxed, reaching the ground state of the system. All calculations were under constant pressure conditions. The final configurations were used for the electronic spectroscopy. We employ the OptaDOS subcode[57,58] for the projected density of states visualization. The geometry optimized supercells are treated as spin-polarized with a grid of 7 x 7 x 7 k-points. The band structure is plotted with an analytical step of 0.001 eV.

**Supporting Information**
Supplementary Figures S1–S19, Supplementary References 1–22

**Acknowledgements**
The research was supported by the Generalitat de Catalunya-AGAUR (2017 SGR 1421) and the European Regional Development Funds (MAT2016-79455-P/FEDER). This project has received funding from the European research Council (ERC) under the European Union's Horizon 2020 research and innovation programme (ULTRASOFC, Grant Agreement number: 681146). Dr. R. Arenal is acknowledged for his support in the TEM analysis. Dr. J. Santiso is acknowledged for his support in the XRD analysis of the epitaxial thin film.

**Author Contributions**
F.C., I.G., A.T., conceived the original idea behind this study. F.C. grew the films, performed the elementary (WDS, EDX), microstructural (SEM, XRD), electrical and electrochemical analysis and wrote the manuscript. I.G. and A.M. supervised and helped with the experimental work and co-wrote the manuscript. L.L.C., G.M., A.R.C., M.W., L.R.G., M.N., S.E., and F.P. carried out the TEM characterization and analysis. A.K. and A.C. contributed with the DFT simulations. A.T. supervised and co-wrote the manuscript.

# Supporting Information

**Engineering Transport in Manganites by Tuning Local Non-Stoichiometry in Grain Boundaries**

*F. Chiabrera, I. Garbayo, L. López-Conesa, G. Martín, A. Ruiz-Caridad, M. Walls, L. Ruiz-González, A. Kordatos, M. Núñez, A. Morata, S. Estradé, A. Chroneos, F. Peiró, A. Tarancón*

F. Chiabrera, Dr. I. Garbayo, Dr. M. Núñez, Dr. A. Morata, Prof. Dr. A. Tarancón
Department of Advanced Materials for Energy, Catalonia Institute for Energy Research (IREC), Jardí de les Dones de Negre 1, Planta 2, 08930 Sant Adrià de Besòs (Barcelona), Spain.

Dr. L. López-Conesa, G. Martín, A. Ruiz-Caridad, Dr. S. Estradé, Prof. Dr. F. Peiró
Department of Electronics and Biomedical Engineering, University of Barcelona, C. de Martí i Franquès 1, 08028 Barcelona, Spain.

Dr. L. López-Conesa, G. Martín, A. Ruiz-Caridad, Dr. S. Estradé, Prof. Dr. F. Peiró
Institute of Nanoscience and Nanotechnology, University of Barcelona, 08028 Barcelona, Spain.

Dr. L. López-Conesa
TEM-MAT Unit, Scientific and Technological Centers of the University of Barcelona (CCiTUB), C. de Lluís Solé i Sabarís 1, 08028 Barcelona, Spain.

M. Walls
Laboratoire de Physique des Solides Bldg. 510, CNRS, Université Paris-Sud, Université Paris-Saclay, Orsay Cedex, 91405 France.

Prof. Dr. L. Ruiz-González
Departamento de Química Inorgánica, Facultad de CC. Químicas, Universidad Complutense de Madrid, 28040 Madrid, Spain.

A. Kordatos, Prof. Dr. A. Chroneos
Faculty of Engineering, Environment and Computing, Coventry University, Priory Street, Coventry CV1 5FB, United Kingdom.

Prof. Dr. A. Chroneos
Department of Materials, Imperial College London, London SW7 2AZ, United Kingdom.

Prof. Dr. A. Tarancón
ICREA, Passeig Lluís Companys 23, 08010 Barcelona, Spain.

E-mail: atarancon@irec.cat

## S.1. Chemical and structural characterization of LSM$_{0.85}$ grain boundaries

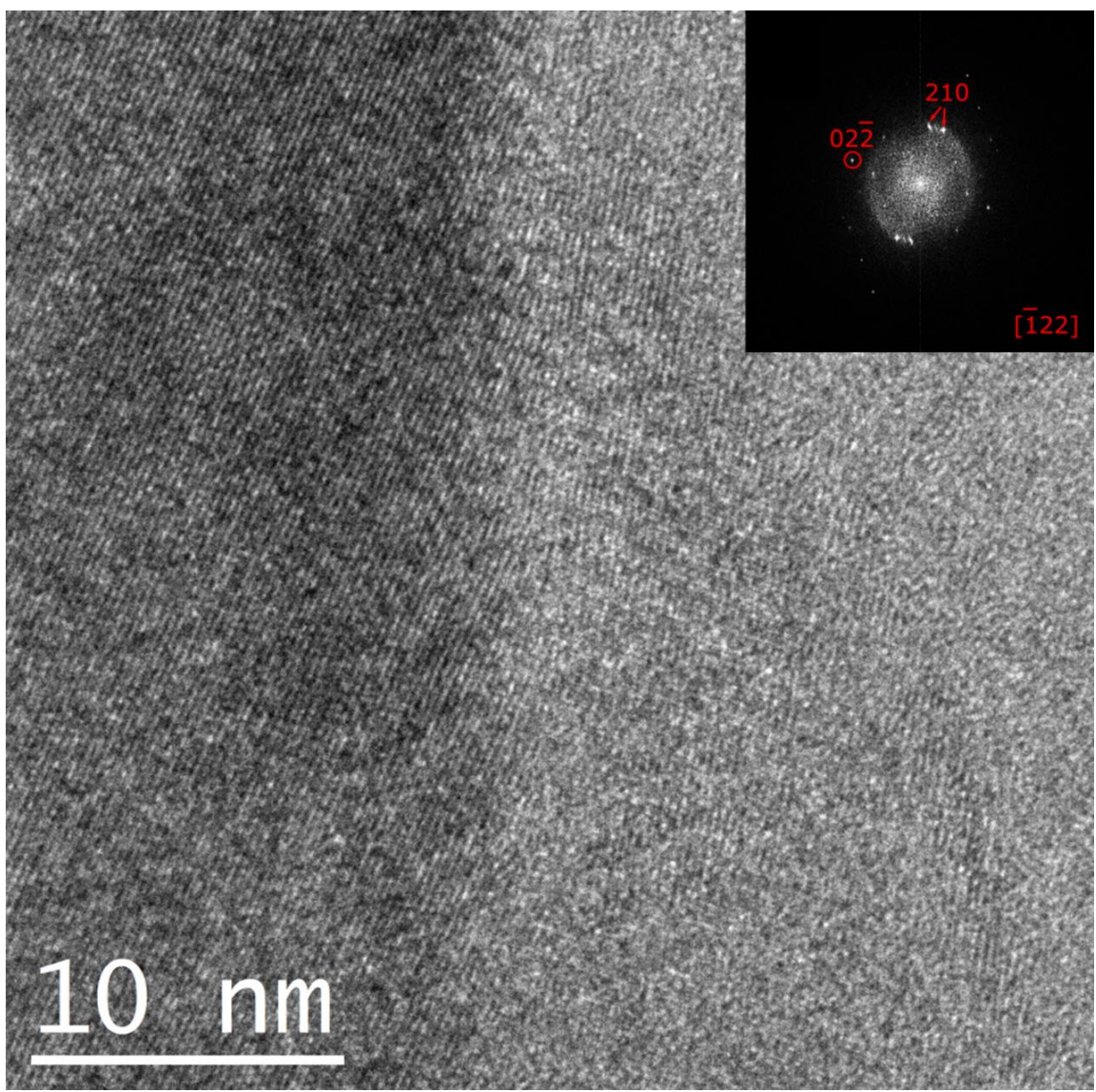

**Figure S1.** HRTEM image containing a grain boundary separating two LSM$_{0.85}$ columnar grains. The inset shows the corresponding FFT, indexed according to the orthorhombic space group 62 (Pnma).

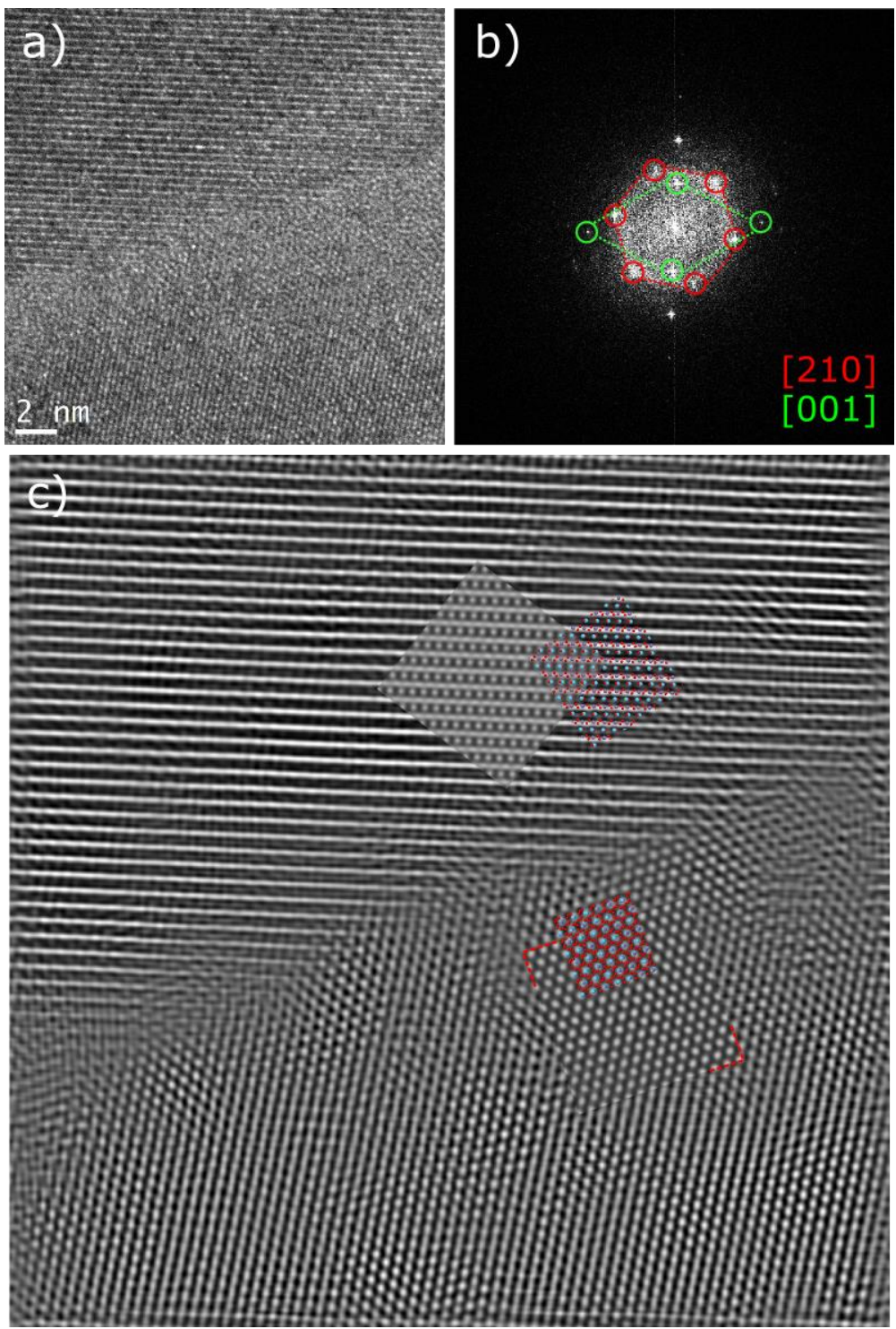

**Figure S2.** HRTEM image containing a grain boundary separating two LSM$_{0.85}$ grains. b) Corresponding FFT, indexed according to the orthorhombic space group 62 (Pnma) for both grains. c) Fourier filtered image. No dislocations are apparent in this grain boundary, although these may be simply not visible in this specific orientation. Insets show atomic models and simulated HRTEM images for both grains.

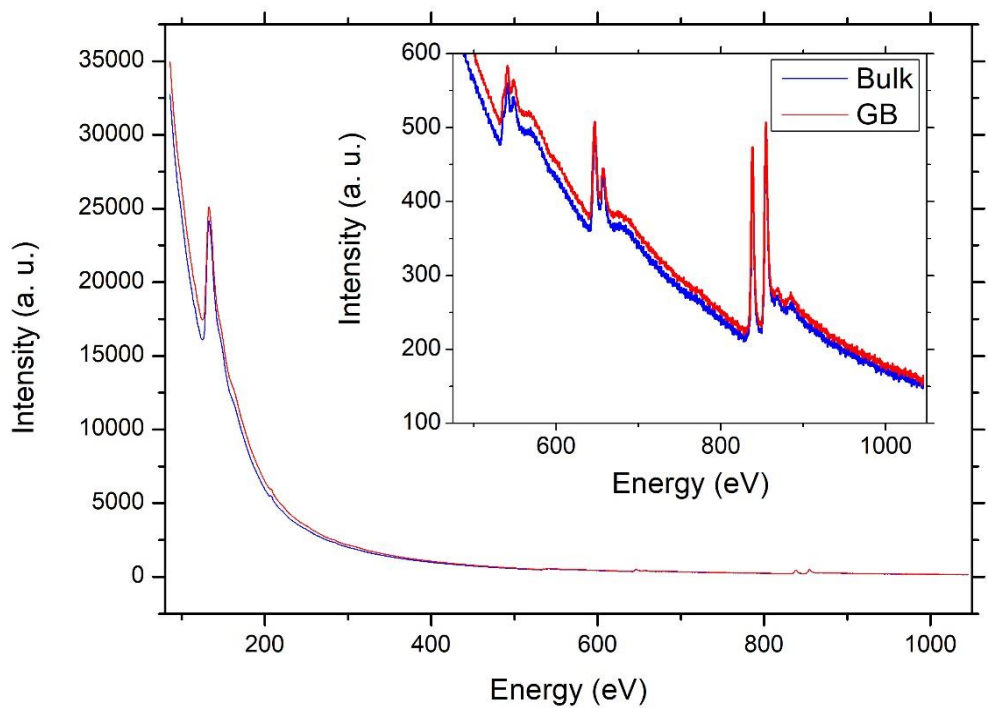

**Figure S3.** EELS spectra of the grain interior and grain boundary region obtained over a large energy window to measure the Sr M edge (133 eV). The inset shows a magnification of the region showing the O K edge (532 eV), Mn L edges (640-651 eV) and La M edges (832-849 eV).

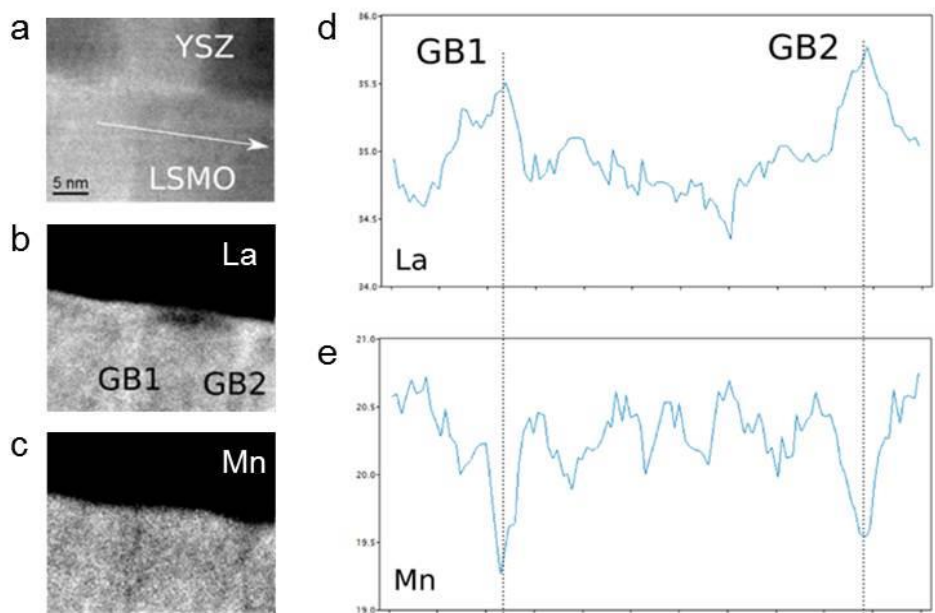

**Figure S4. a.** HAADF figure of two grains near the interface with YSZ film. EELS compositional maps of La (**b**) and Mn (**c**) concentrations. Compositional variation of La (**d**) and Mn (**e**) along the white arrow drawn in (**a**). The presence of La enrichment and Mn deficiency in two consecutive GBs shows the uniformity of the cationic variations along the film.

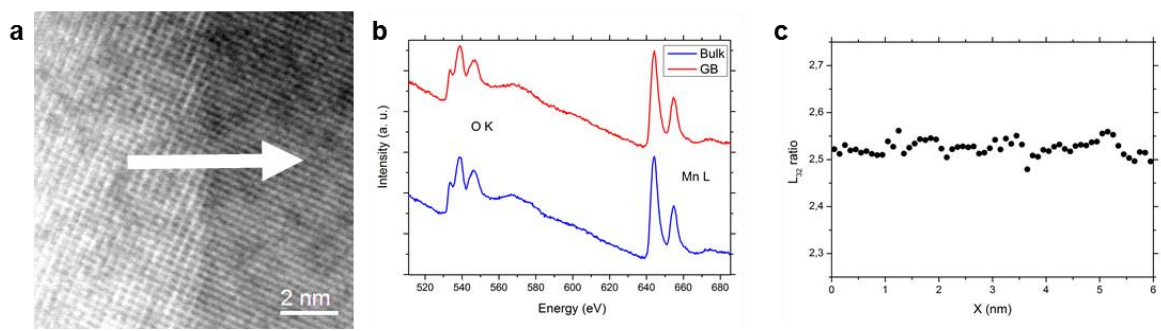

**Figure S5. a.** HADDF of a grain boundary. **b**. EELS of grain and grain boundary regions obtained over a small energy window in order to detect variance of oxidation state. **c**. Mn $L_{32}$ ratio across a grain boundary showing a constant profile and indicating no Mn valence change.

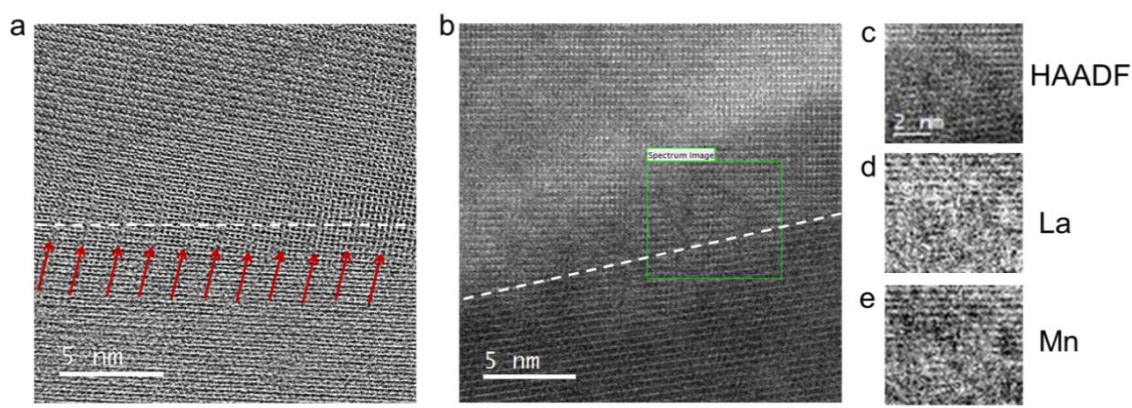

**Figure S6. a,** STEM bright field image of the grain boundary analyzed by EELS along the interface. The red arrows highlight the atomic disordered fringes at the GB interface created by the arrays of dislocations. **b,** STEM-HAADF image of the same GB depicted in **a. ** The white dashed line indicates the GB position in both the TEM figures **a** and **b. c**, STEM-HAADF image of the green square shown in **b**. **d** and **e** shows the La and Mn compositional maps extracted by EELS analysis.

## S.2. Characterization of epitaxial LSM$_{0.85}$ on NdGaO$_3$ (NGO) (110) substrate

The epitaxial thin film of LSM$_{0.85}$ on NdGaO$_3$ (110)$_o$ (where o stands for orthorhombic) substrate was analysed by XRD (**Figure S7**). The LSM film shows a unique orientation, corresponding to the (00l)$_{pc}$ (where pc stands for pseudo-cubic). **Figure S7b** shows the high resolution XRD of the LSM$_{0.85}$/NGO structure around the (002)$_{pc}$ peak. The film is highly oriented and the finite-size oscillations demonstrate the long-range homogeneity of the film. A thickness of 60 nm is extracted from the fitting of the oscillations, consistent with the thickness measured by ellipsometry. The phi scan around the 220 peak (inset of **Figure S7b**) and the reciprocal space maps around the $\bar{3}$03 and the 033 asymmetrical peaks (**Figure S8**) show that the film is coherently strained by the substrate. Therefore, a "cube on cube" growth of the film on the substrate is expected[1] (schematically shown in **Figure S7c**). The out of plane lattice parameter obtained for the LSM$_{0.85}$ thin films is c$_{pc}$=3.96 Å, which is higher than the one expected from the compressive strain imposed by the substrate (-0.70% and -0.47% on a$_{pc}$ and b$_{pc}$ respectively). Such increase of the out of plane lattice parameter is likely derived from the Mn deficiency measured by Wavelength Dispersive Spectroscopy (WDS), and goes in line with the results obtained in other B site deficient perovskites.[2–4]

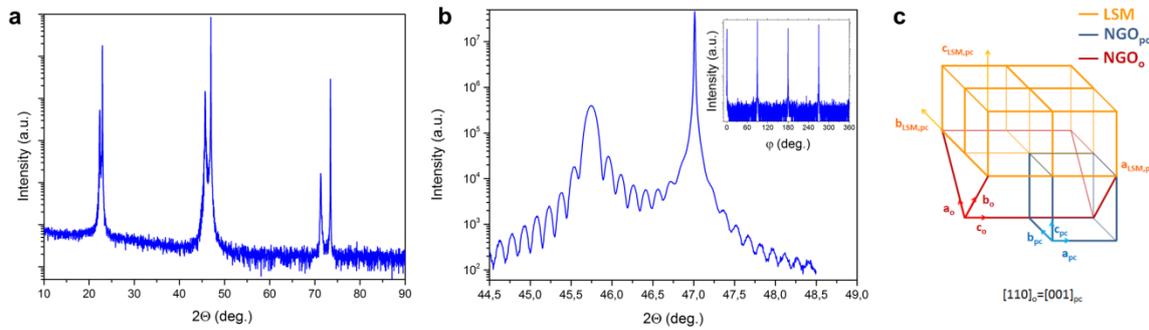

**Figure S7. a,** XRD of the LSM$_{0.85}$ thin film grown on NGO (110)$_O$ substrate. **b**, High resolution XRD around the 200 peaks showing finite-size oscillations originating from the long-range homogeneity of the film. The inset shows the φ scan around the 220 peak. **c,** scheme of the epitaxial relation between pseudo-cubic LSM and orthorhombic NGO (110)$_o$.

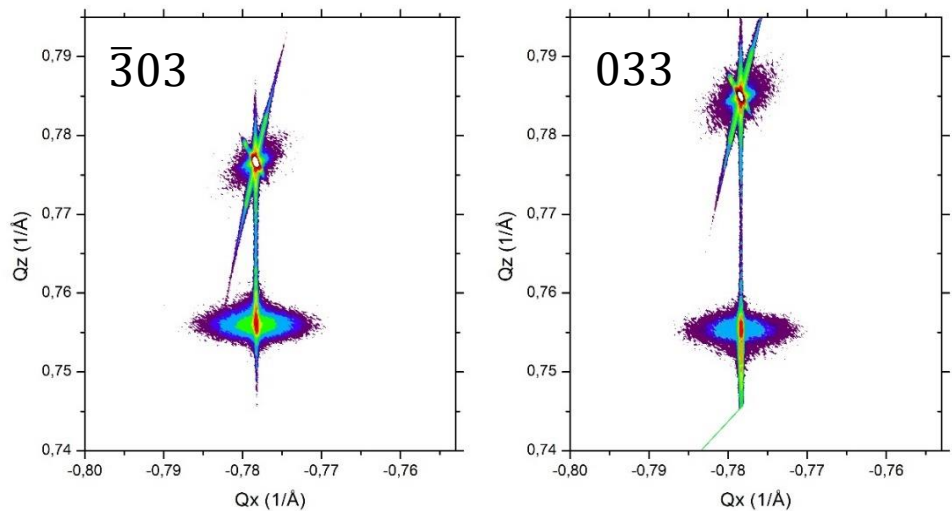

**Figure S8.** Reciprocal space maps of $\bar{3}$03 and 033 in plane reflection showing that the film is coherently strained by the substrate.

## S.3. Grain boundary engineering through the Mn/(La+Sr) thin film variation

### S.3.1. Description of the Combinatorial Pulsed Laser Deposition (C-PLD) approach

In this work, Combinatorial Pulsed Laser Deposition (C-PLD) was used to grow $La_{0.8}Sr_{0.2}Mn_{1-y}O_{3\pm\delta}$ ($LSM_y$) with variable content of Mn. C-PLD is a technique based on the combinatorial deposition of two parent compound (in this work LSM and $Mn_3O_4$) on a large area substrate.[5–8] **Figure S9a** shows the schematic representation of the technique. The deposition sequence consists in the alternate deposition of the two materials plumes focused at a relative distance $d$ between them. Taking advantage of the gaussian-shape of the plasma plume, a gradient of concentration is developed between the plumes centres, determining in one single deposition all the possible combinations between the two materials (**Figure S9b**). The thickness deposited per cycle and the deposition temperature must be carefully chosen in order to allow an in-situ synthesis of the two materials. In this work, a substrate temperature of $T = 700ºC$ and a thickness of 1 nm and 0.25 nm per-cycle for LSM and $Mn_3O_4$ respectively are selected to assure a proper cation diffusion.[9,10] The plume distance was set at $d = 8$ cm to obtain a gradual increase of Mn content in the films. Different substrates were chosen for performing electrical and electrochemical experiments. For the former, Si (100) 4" wafer pre-passivated by 100 nm YSZ and 5mm x 10mm Sapphire (0001) single crystal substrates were used. For the out of plane electrochemical experiments, 3x3 mm YSZ (001) single crystal substrates were preferred for their pure oxygen ionic conduction behaviour. Sapphire and YSZ substrates were attached with Silver paste to a 4" Si wafer forming a line between the two plume centres, allowing a gradual increase of B/A=Mn/(La+Sr) ratio.

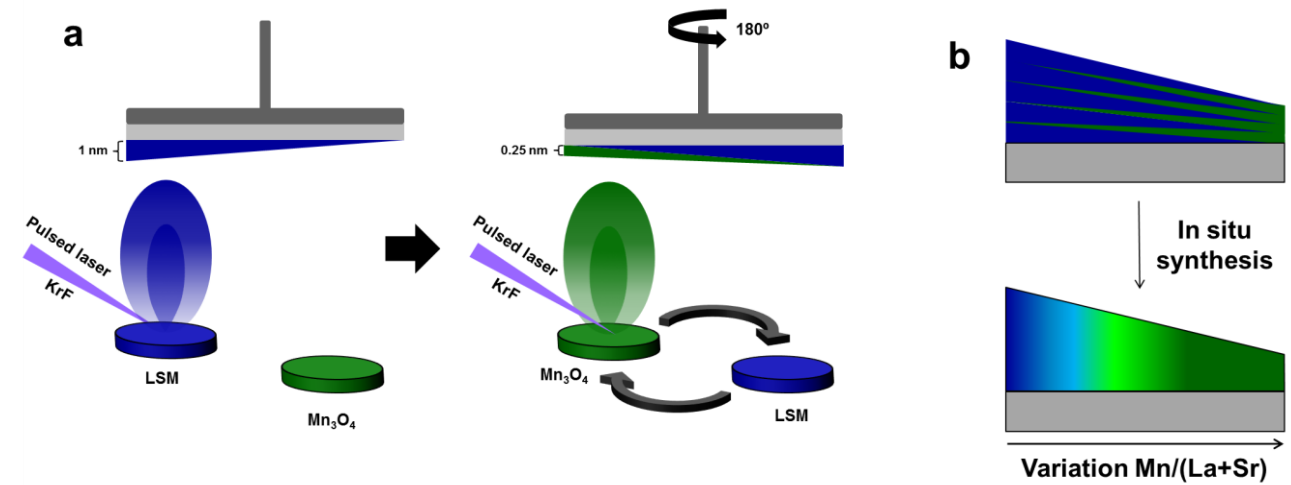

**Figure S9. a.** Sketch of the C-PLD approach employed for the deposition of $LSM_y$ films with different B/A ratio. **b**. Concentration gradient developing between the two plume centres due to high temperature in-situ synthesis of the two parent compounds.

### S.3.2. Derivation of B/A ratio and structural characterization of the combinatorial $LSM_y$

First, the parent compounds plumes were studied. **Figure S10a** and **b** shows respectively the thickness of LSM and $Mn_3O_4$ deposited on Si substrates, obtained by ellipsometry. The plumes were fitted with a Pseudo-Voigt curve.

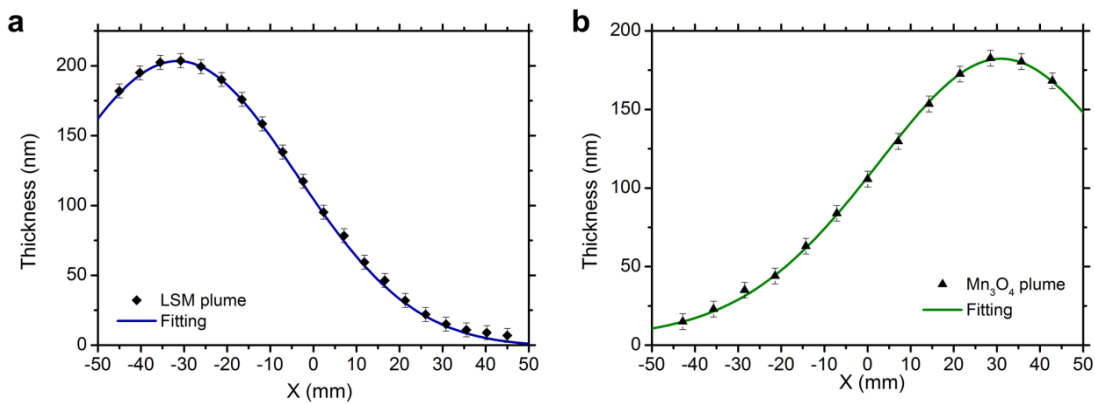

**Figure S10.** Thickness measured by ellipsometry for a. LSM and b. $Mn_3O_4$ plumes. The curves were fitted using a Pseudo-Voigt function.

The $Mn_3O_4$ plume was analysed by XRD and SEM and Raman spectroscopy (**Figure S11**), revealing that the film has the same phase of the target.

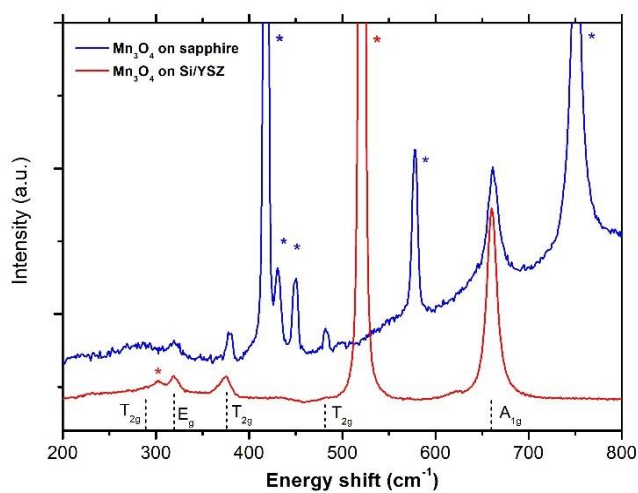

**Figure S11.** Raman spectra of the $Mn_3O_4$ deposited on sapphire (blue line) and YSZ/Si (red line). Both samples present the characteristic bands of the $Mn_3O_4$ phase. In the graph, blue and red asterisk shows respectively the sapphire and Si peaks.

**Figure S12a** shows the thickness profile measured by ellipsometry along the plumes centres for the combinatorial sample $LSM_y$ deposited on YSZ/Si. The data were fitted by a superimposition of the parent compound plumes obtained before, leaving as free parameters only the plume maxima and position centres. The fitting shows a good similarity with the experimental data. Once the effective thickness and the exact position of the parent compounds plumes is known, the predicted thickness can be calculated as:

$$\frac{Mn}{La+Sr} = \frac{moles\ Mn_{LSM} + moles\ Mn_{Mn_3O_4}}{moles\ La_{LSm} + moles\ Sr_{LSM}} \qquad \text{Equation S1}$$

Transforming the equation, we get:

$$\frac{Mn}{La+Sr} = \frac{1 \cdot \rho_{LSM} \cdot t_{LSM}/M(LSM) + 3 \cdot \rho_{Mn_3O_4} \cdot t_{Mn_3O_4}/M(Mn_3O_4)}{(0.8+0.2) \cdot \rho_{LSM} \cdot t_{LSM}/M(LSM)} \qquad \text{Equation S2}$$

Where $\rho$ is the density, M the molar mass, t the thickness. The result as a function of wafer position is shown in **figure S12b** along with the B/A ratio measured by Energy Dispersive X-ray Analysis (EDX). It is clearly visible that the two methods yield to the exact same B/A ratio. In this study, a composition span from B/A = 0.85 to 1.2 (corresponding to X = -45 mm and -8 mm, respectively) was analysed. For thin films with composition out of this range secondary phases were observed.

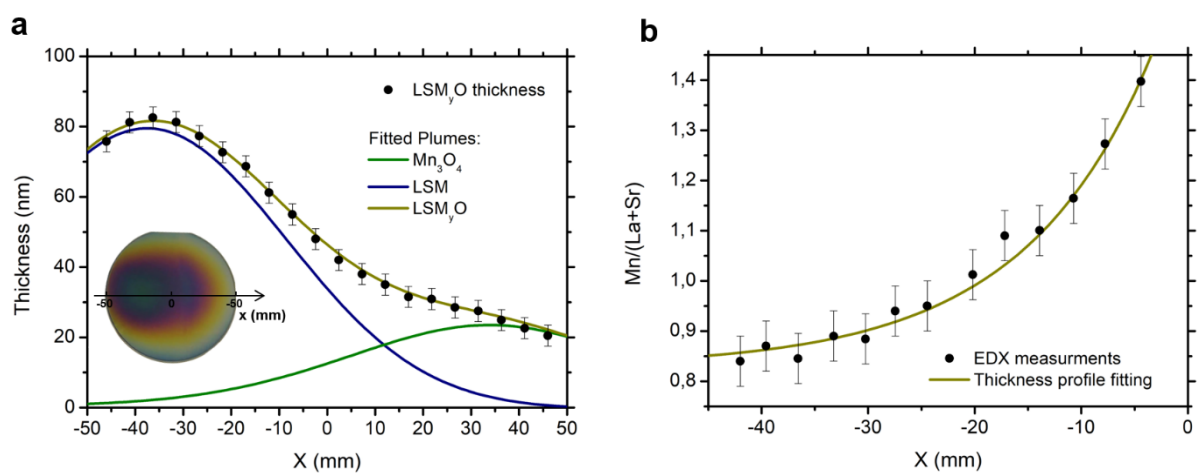

**Figure S12. a.** Thickness of the LSM$_y$ combinatorial sample measured by ellipsometry and fitted by the superimposition of the parent compounds plumes. **b.** B/A ratio measured by EDX and calculated by equation S2 using the single plumes thickness obtained in a.

The surface of the LSM$_y$ sample was studied by scanning electron microscopy (SEM). **Figure S13** shows the top-view of the LSM$_y$ for different B/A ratio concentration. No substantial evolution of the grains is observed, meaning that the fraction of the GB area remains nearly constant. This point is particularly important for the electrochemical analysis, since a modification of the grain size would result in a modification of oxygen mass transport properties.[11]

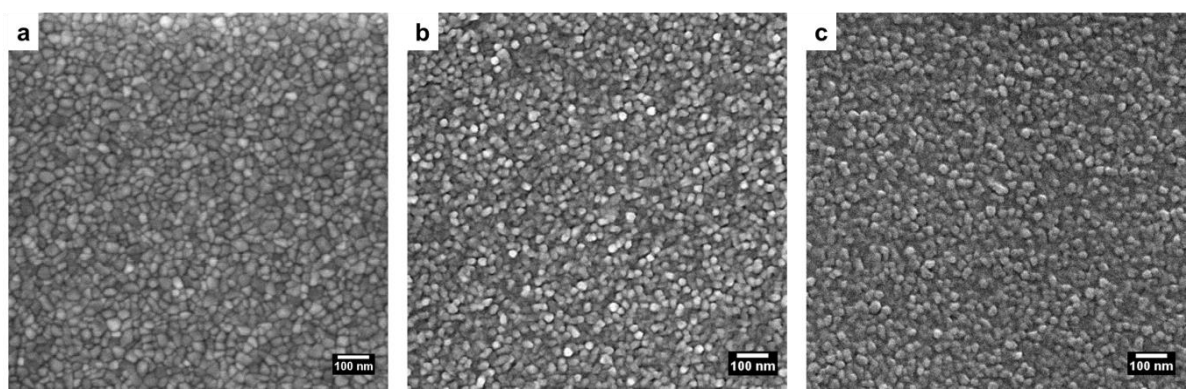

**Figure S13.** SEM top view of LSM$_y$ samples deposited on YSZ/Si for **a.** B/A=0.92, **b.** B/A=0.97 and **c.** B/A=1.05. No significant microstructural evolution is observed.

**Figure S14** shows the variation of oxygen content obtained by EELS scan across three GBs with different Mn content (the GBs are the same reported in **Figure 3** of the main text). For all the composition analysed, a decrease of oxygen concentration is measured at the interface.

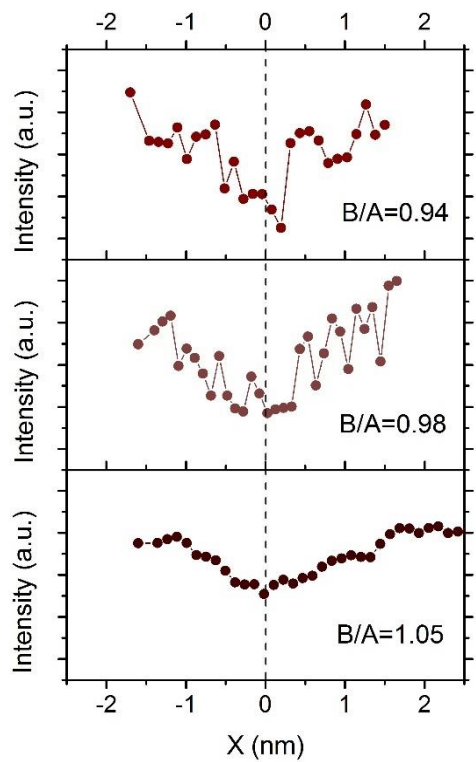

**Figure S14.** Compositional variation of O obtained by EELS in three GBs of LSMy samples deposited on YSZ/Si for B/A=0.94, B/A=0.98 and B/A=1.05. The GBs analysed are the very same of **Figure 3** of the main text.

*S.3.3. Discussion on the quantification of cationic composition in LSM$_y$ grain boundaries*

**Figure S15a** and **b** shows the Mn and La variation measured across the GB by EELS for LSM$_y$ samples with different B/A ratio. The cation concentration was quantified knowing the concentration in the grain bulk, taken from EDX measurements (see Figure S12b). The figure shows that the Mn-deficient GB (B/A = 0.85) present the highest cationic change, while in the Mn rich GBs the absolute change tends to decrease while increasing the B/A ratio. In particular, in the Mn deficient GB, the Mn concentration in the GB is extremely low (around 0.75). Since no secondary phases were detected up to the GB core, this significant Mn deficiency, along with the La increase, strongly suggests the formation of antisite defects La$_{Mn}$ in order to maintain the perovskite structure.

It is important to note, though, that the quantification of GB composition in polycrystalline thin films can be susceptible to a non-negligible error. This is because slight differences in the GB angles with respect to the TEM incident beam can partially mask the compositional changes. The tilt of the GB angle can therefore explain differences in the ionic composition when comparing different GBs from a sample with similar overall cation content, such as the interfaces for B/A = 0.85 analysed here and the ones of **Figure 2b** of the manuscript. Nevertheless, the qualitative trends presented in this work were fully consistent in the entire set of GB analyzed for each composition.

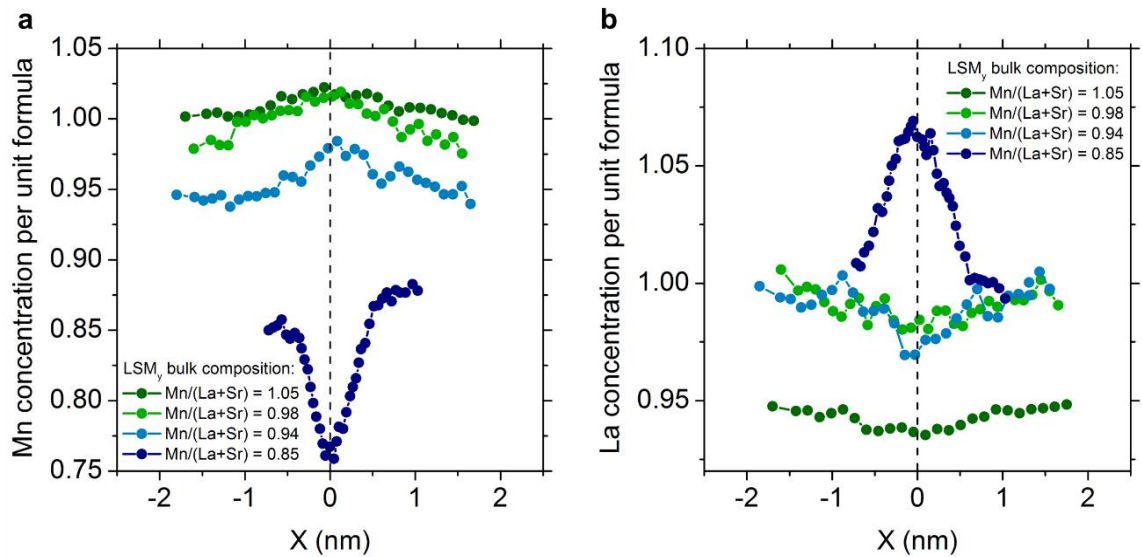

**Figure S15.** Quantitative compositional variation of Mn (**a**) and La (**b**) obtained by EELS in the GBs of LSMy samples deposited on YSZ/Si for B/A=0.85, B/A=0.94, B/A=0.98 and B/A=1.05.

### *S.3.4. In plane electrical properties of LSM$_y$*

*S.3.4.1. 4-probe in-plane measuring configuration*

The LSM$_y$ combinatorial samples were analyzed in an in-plane four probe geometry, in accordance with our previous work.[12] 1x0.5 cm$^2$ rectangular Sapphire substrates were acquired from Crystal GmbH (Germany), while the YSZ/Si substrates were prepared in house from a 4" wafer cut into chips of 1x0.5 cm$^2$. **Figure S16** shows a schematic representation of the measuring system.

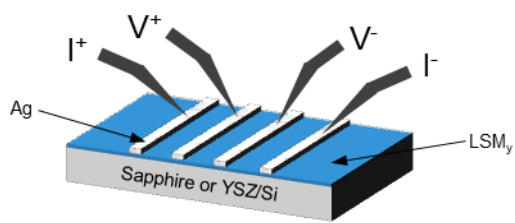

**Figure S16.** Schematic representation of the 4-probe in-plane method used for electrical characterization of the LSM$_y$ combinatorial sample.

The distance between the two central electrodes (designed to reach both sides of the chips in order to minimize the current leakage) was set to be 1mm. The variation of Mn concentration occurring inside the voltage electrodes was considered in the compositional error (error bars in **Figure 4c**). The resistance of the thin films was analyzed in a linear regime.

*S.3.4.2. Electrical resistivity of LSM$_y$ thin films on different substrates*

**Figure S17** shows the in-plane electrical resistivity of the LSM$_{0.85}$ measured as deposited and after a long annealing (5 hours) at 650ºC. No significant changes were observed.

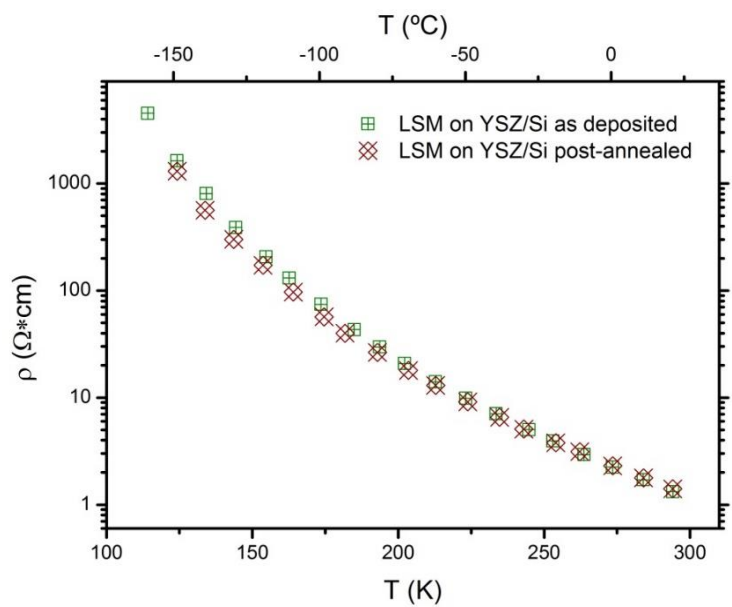

**Figure S17.** Electrical resistivity as a function of temperature of Mn-deficient polycrystalline LSM$_{0.85}$ thin films as deposited and annealed at 650ºC for 5 hours.

**Figure S18** shows the in-plane electrical resistivity measured for LSM$_y$ films deposited on YSZ/Si and Sapphire. The resistivity shows no difference between the two substrates, while the B/A ratio appears to have a considerable influence on the electrical properties. For B/A > 0.92 the system starts to recover the M-I transition, which then progressively takes place at higher temperature increasing the Mn content. This behaviour is ascribable to the local defect concentration at the grain boundaries, since the epitaxial sample with B/A = 0.85 normally displays the M-I transition.

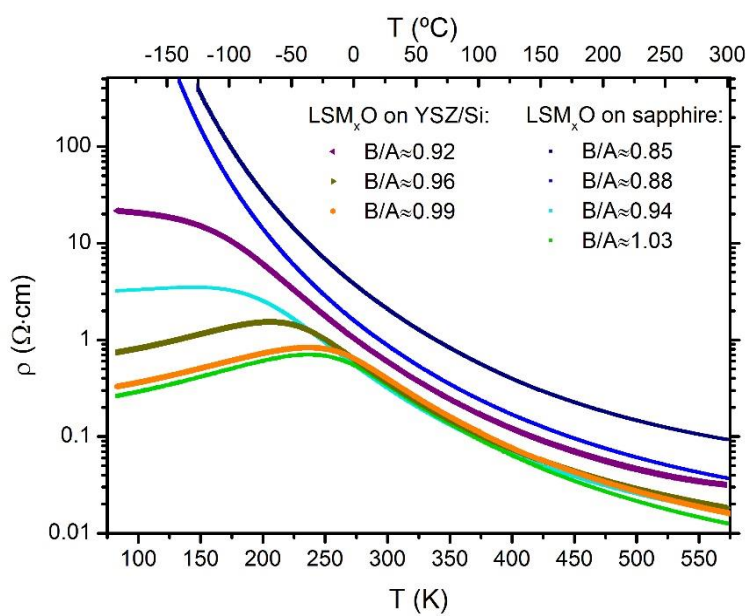

**Figure S18.** In-plane electrical resistivity as a function of temperature measured for the LSM$_y$ deposited on YSZ/Si and sapphire.

We investigated the effect of varying the B/A ratio on the bulk electrical properties by studying three epitaxial thin films with composition B/A = 0.85, B/A = 1.03, B/A = 1.2. **Figure S19** shows the in-plane electrical resistivity measured for these thin films. All the epitaxial thin films show the expected M-I transition at low temperature. An increase of M-I temperature is observed increasing the B/A ratio (see arrows in the figure), probably due to the detrimental effect of random Mn vacancies on the metallic order. In the Mn-rich film, a further small improvement of the M-I transition temperature is observed.

These bulk changes are expected to play a secondary role on the conductivity of the polycrystalline samples both for low and high Mn content, since these films present orders of magnitude higher resistivity respect to the epitaxial thin films. The GBs are therefore expected to dominate the in-plane resistivity of the polycrystalline $LSM_y$ thin films.

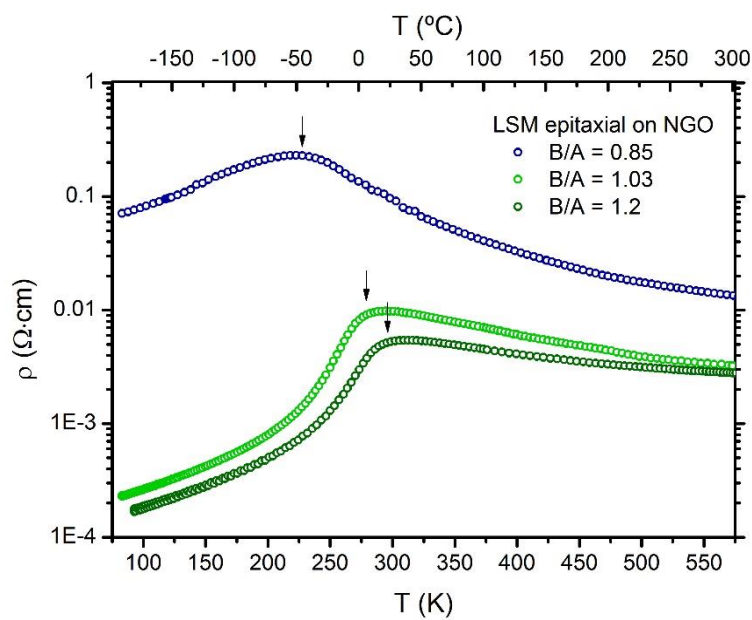

**Figure S19.** In-plane electrical resistivity as a function of temperature measured for the epitaxial $LSM_y$ with different B/A ratios, deposited on NGO. The arrows indicate the M-I transition.

### S.3.5. Out of plane electrochemical properties of $LSM_y$ on YSZ single crystal

Electrochemical impedance spectroscopy (EIS) is a powerful tool for characterizing the oxygen mass transport properties of high temperature MIECs.[13] In order to be able to distinguish the single phenomena affecting the overall oxygen electrochemical properties, MIECs thin films need to be deposited on a pure oxygen conductor. Hence, Combinatorial $LSM_y$ was deposited on 3 mm x 3 mm YSZ (001) single crystals. The small dimensions of the crystals allowed minimizing the compositional changes inside the single samples. Silver paste was used as counter back electrodes due to its high electrochemical properties and it is expected to give a negligible effect in the electrochemical resistance.[11] Porous gold paste was used on the LSM for permit a fast in plane electronic conduction, being gold known to be mostly inactive for ORR reactions.[14] In this way, porous gold / $LSM_y$ / YSZ / porous silver cells were characterized by EIS at 650 ºC in synthetic air. The temperature was set lower than the deposition one to minimize the grain growth during the measurements. **Figure S20b** and **S20c** shows respectively the Nyquist and Bode plot obtained for the B/A ≈ 0.89 sample. Two major contributions are clearly visible, corresponding to high and low frequency. The spectrum well agrees with the typical MIEC behavior, where the two distinguishable phenomena correspond to the oxygen diffusion and surface reactions, in agreement with our previous work[15] and with literature.[11,16] In order to derive the resistance associated with these two contributions, the spectrum has been fitted using the characteristic MIEC equivalent circuits (**Figure S20a**), first described by Jamnik and Maier[17] and widely used for LSM.[11,15,16] The circuit includes a ZARC element accounting for the surface oxygen incorporation processes and a transmission line with two different pathways: the electronic one (without resistance associated) and the ionic one (with infinitesimal resistances $dR_i$). The two rails are connected by a series of infinitesimal capacitances $dC_{chem}$, which describes the chemical energy stored in a material under the application of an electrical bias.[17] Finally, a pure capacitance is placed at the end of the electronic rail, describing the blocking electrons surface between LSM and YSZ. As can be seen in **figure S20**, the circuit perfectly reproduces the $LSM_y$ behavior. All the impedance spectra measured for the $LSM_y$ samples well agrees with the MIEC equivalent circuit. A clear evolution of the overall electrochemical resistance is observed changing the B/A ratio (**Figure S21**).

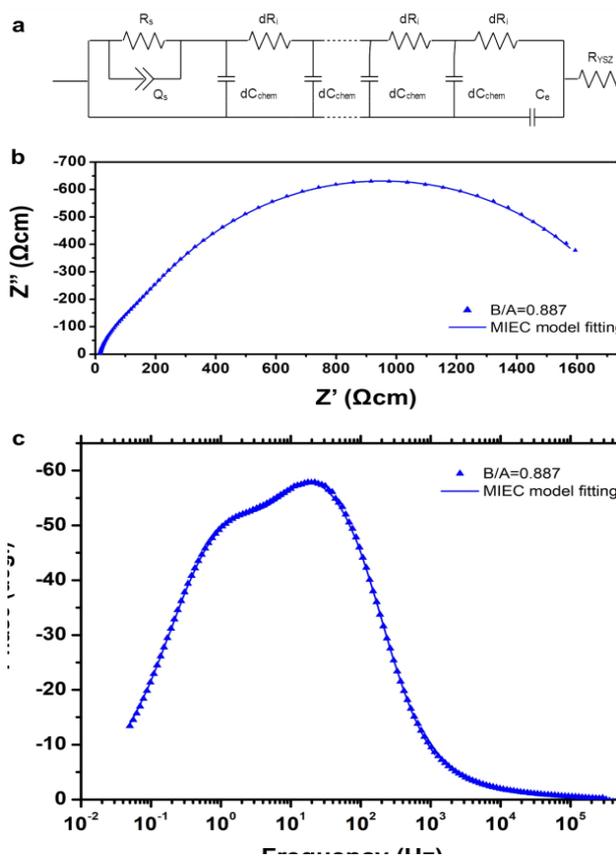

**Figure S20. a.** Equivalent circuit employed for the fitting of the $LSM_y$ impedance. **b.** Nyquist and **c.** Bode plot of the impedance spectra obtained at 650 °C in synthetic air for the B/A = 0.89 sample.

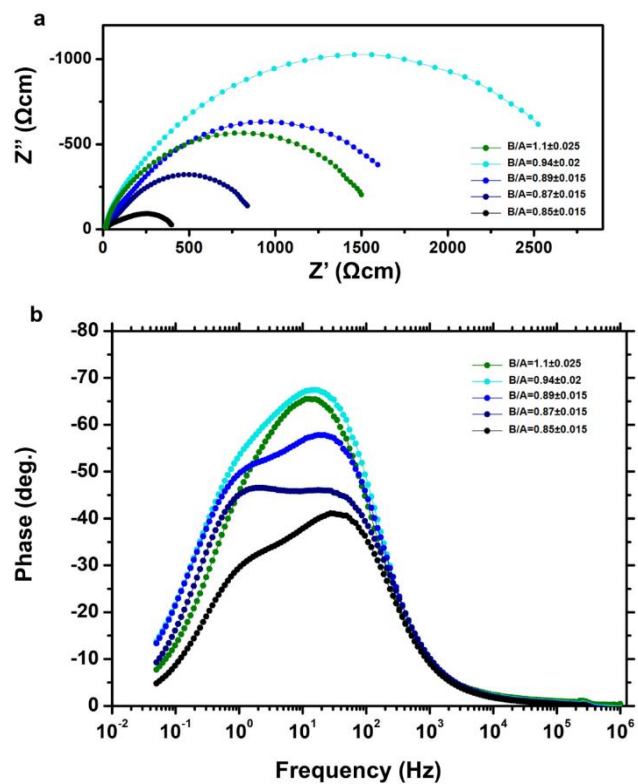

**Figure S21. a.** Nyquist and **b.** Bode plots of the impedance spectra obtained at 650 ºC in synthetic air for different B/A ratio samples. A clear evolution of both the diffusion and surface exchange phenomena is observed.

The ionic and surface resistance obtained in the impedance fitting can be used to derive the overall surface exchange coefficient $k^q_{av}$ and the oxygen diffusion $D^q_{av}$. The diffusion coefficient is calculated by:

$$D^q_{av} = \frac{L}{R_{ion} \cdot A} \cdot \frac{k_b T}{c_{O2} z_i^2 e^2} \qquad \text{Equation S3}$$

Where $R_{ion}$ is the ionic resistance, $k_b$ is the Boltzmann constant, $z_i$ is the number of charges involved in the transport, $L$ and $A$ are respectively the thickness and the surface area of the LSM, $e$ is the elementary charge and $c_{O2}$ is the concentration of oxygen in the LSM lattice. The surface exchange coefficient can be calculated from the surface resistance ($R_s$) by:

$$k^q_{av} = \frac{1}{R_s \cdot A} \cdot \frac{k_b T}{z_i^2 e^2 c_{O2}} \qquad \text{Equation S4}$$

The coefficient $D^q$ obtained by electrical measurements can be confronted with $D^*$ (obtained in the oxygen tracer experiments) considering a correlation factor of $f = 0.69$.[18,19] Also the confrontation between $k^q$ and $k^*$ is possible.[18]

It is important to notice that the diffusion and surface exchange parameters obtained in this study represents an averaged value between the grain and grain boundary, since EIS is unable to distinguish the two contributions which take place in parallel. This points out that in order to permit a reasonable confrontation with the coefficients extracted in SIMS experiments[20–22] (in which grain and grain boundary coefficients were separated) one has to average $D^*$ and $k^*$ with the effective area that grain and grain boundary regions occupy. Being the structure of these thin films columnar, it is straightforward to calculate the overall D and k that these films would display in an electrochemical experiment. The dimensions of grain and grain boundary used for the calculation are the ones used for the SIMS oxygen concentration fitting in the different works.[20–22]